**Electrically driven spin resonance in silicon carbide color centers**

P. V. Klimov[1,2], A. L. Falk[1,2], B. B. Buckley[2], and D. D. Awschalom[1,2]

1. Institute for Molecular Engineering, University of Chicago, Chicago, IL 60637, USA

2. Center for Spintronics and Quantum Computation, University of California, Santa Barbara, CA 93106, USA

**We demonstrate that the spin of optically addressable point defects can be coherently driven with AC electric fields. Based on magnetic-dipole forbidden spin transitions, this scheme enables spatially confined spin control, the imaging of high-frequency electric fields, and the characterization of defect spin multiplicity. While we control defects in SiC, these methods apply to spin systems in many semiconductors, including the nitrogen-vacancy center in diamond. Electrically driven spin resonance offers a viable route towards scalable quantum control of electron spins in a dense array.**

PACS number(s): 42.50.Ex, 71.70.Ej, 76.30.Mi, 76.30.-v

Optically addressable paramagnetic defects have proven to be powerful systems for solid-state quantum control. Research into the nitrogen-vacancy (NV) center in diamond has been driven by applications in quantum information and nanoscale sensing [1-7]. More recently, intrinsic defects in SiC [8] have been shown to have similar properties to the NV center in diamond, including long coherence times that persist up to room temperature [9], a high degree of optical polarization [10], and spin-dependent photoluminescence [9-15]. The technological maturity of SiC growth and processing combined with defect emissions near telecom wavelengths make SiC defects particularly amenable to integration with electronic, optoelectronic, electromechanical, and photonic devices.

An important challenge in defect spin physics is to selectively manipulate individual spins at the nanometer scale. Localized spin control is particularly important since the inter-spin separation required for strong dipolar coupling is on the order of tens of nanometers [16, 17]. Since electric fields are readily confined on similar length scales [18], electrically driven spin resonance [19-25] could be used to address this challenge. In this letter, we show that resonant electric fields can coherently drive spin transitions in optically addressable defects.

We use AC electric fields to drive Rabi oscillations across a magnetic-dipole forbidden spin transition ($\Delta m_s=\pm 2$) of the optically addressable electronic spin of the QL1 defect in semi-insulating 6H-SiC. We then apply our electrically driven, optically detected magnetic resonance (E-ODMR) technique to spatially map the QL1 spin response to an AC electric field generated by our fabricated electrodes. This imaging applies to GHz-frequency resonant electric fields, complementing non-resonant kHz-frequency AC electric field sensing previously demonstrated with NV centers in diamond [3]. Since the QL1 defect shares a ground-state spin Hamiltonian with many intrinsic defects in SiC [8-10, 13, 15, 26-29] and the NV center in diamond [4], our results apply to a broad class of optically addressable solid-state defects.

Our E-ODMR measurements determine that the optically addressable spin of QL1 has integer-value total spin ($S$), and long spin relaxation ($T_1$) times [10] suggest that it is the orbital ground state (see SI). Together with its similar optical and spin transition energies to the $S$=1 neutral divacancies in 4H-SiC [8, 9], QL1 is likely to be an $S$=1 neutral divacancy as well. Its c-axis orientation and $C_{3v}$ point-group symmetry lead to the ground-state spin Hamiltonian [10, 30]:

$$H = (hD + d_\parallel E_z)\sigma_z^2 + g\mu_B \boldsymbol{\sigma} \cdot \boldsymbol{B} - d_\perp E_x(\sigma_x^2 - \sigma_y^2) + d_\perp E_y(\sigma_x\sigma_y + \sigma_y\sigma_x), \qquad (1)$$

where the c-axis is oriented along the z-axis, $h$ is Planck's constant, $\mu_B$ is the Bohr magneton, $D$ = 1.299 GHz [10] is the zero-field splitting, $g$ = 2 is the g-factor, $\boldsymbol{\sigma}$ is the vector of spin-1 Pauli matrices (see Supplemental Information (SI)), $\boldsymbol{B}$ is the magnetic field vector, and $\boldsymbol{E}$ is the electric field vector. The longitudinal (c-axis oriented) and transverse coefficients $d_\parallel$ and $d_\perp$, respectively, couple electric fields to the spin via the Stark effect. The Zeeman-split energy levels due to a longitudinal magnetic field ($B_\parallel$) are illustrated in Fig. 1(a).

The electrically-driven spin resonance that we demonstrate is similar to standard magnetically-driven spin resonance, except that it couples different pairs of spin eigenstates. When written in the $\sigma_z$ basis, our energy eigenstate basis, the transverse magnetic and electric components of Eq. 1 can be written as:

$$H_\perp^B = g\mu_B(B_x\sigma_x + B_y\sigma_y) = \frac{g\mu_B}{\sqrt{2}}B_\perp e^{-i\phi_B}(|+1\rangle\langle 0| + |0\rangle\langle -1|) + H.c., \qquad (2)$$

$$H_\perp^E = -d_\perp E_x(\sigma_x^2 - \sigma_y^2) + d_\perp E_y(\sigma_x\sigma_y + \sigma_y\sigma_x) = -d_\perp E_\perp e^{i\phi_E}|+1\rangle\langle -1| + H.c., \qquad (3)$$

where $|i\rangle$ is defined to be $|m_s = i\rangle$, $B_\perp(E_\perp)$ and $\varphi_{B,(E)}$ are the magnitude and phase, respectively, of the magnetic (electric) field in the plane transverse to the c-axis, and h.c. denotes the Hermitian conjugate. The main difference between $H^B_\perp$ and $H^E_\perp$ is that $H^B_\perp$ connects triplet pairs with $\Delta m_s$=±1, whereas $H^E_\perp$ connects triplet pairs with $\Delta m_s$=±2. As such, in the same way that applying resonant transverse magnetic fields can be used to drive magnetic-dipole ($\Delta m_s$=±1) transitions, resonant transverse electric fields can be used to drive magnetic-dipole forbidden ($\Delta m_s$=±2) transitions.

Our experiments use both AC electric and magnetic field control, for which we use separate driving elements. Open-circuit interdigitated metal electrodes on the chip's top surface are used to drive transverse electric fields between adjacent digits [Fig. 1(b)], and a short-circuited stripline beneath the chip is used to drive transverse magnetic fields over the interdigitated region. A permanent magnet provides a static $B_\parallel$. QL1 color centers were produced in our 6H-SiC substrate via a carbon implantation and annealing process designed to generate defects in a 400 nm thick layer immediately below the surface (see SI). The QL1 spins between adjacent electrode digits are optically addressed by non-resonantly pumping their 1.09 eV near-infrared optical transition with 1.27 eV laser light (see Fig. 1(c) and SI for details).

Much like the NV center in diamond [30], the QL1 spin-dependent optical cycle allows non-resonant laser illumination to both polarize and read out its ground-state spin. Because the defects' photoluminescence intensity ($I_{PL}$) depends on whether QL1 is in the $|0\rangle$ or the $|\pm1\rangle$ spin states, we can track the spin dynamics by measuring differential fluorescence ($\Delta I_{PL}$) between an initial state and one that has been evolved by magnetic or electric field pulses. These $\Delta I_{PL}$ measurements thus enable conventional (magnetically driven) optically detected magnetic resonance (ODMR) [Fig. 1(d)] and E-ODMR (see SI).

We measure E-ODMR by implementing the sequence of spin transitions shown in Fig. 2(a) [31]. We first optically initialize the spin ensemble into $|0\rangle$ (by convention, see SI) and then rotate it into $|-1\rangle$ with a magnetic π-pulse driven on the stripline. We then generate a microwave-frequency pulse (P) on the interdigitated electrodes to transfer population between $|-1\rangle$ and $|+1\rangle$ using electrically driven spin resonance. Any spin population remaining in $|-1\rangle$ is transferred back into $|0\rangle$ with another magnetic π-pulse driven on the stripline. After this sequence we re-illuminate the sample to read out the ensemble magnetization and to re-initialize it. By modulating P on and off and measuring the locked-in $\Delta I_{PL}$ signal, we have a direct measurement of $\Delta m_s=\pm2$ transitions driven by P.

To observe E-ODMR in the frequency domain, we fix the length of P and sweep its frequency [Fig. 2(b)]. This experiment returns a clear resonance at exactly the $\Delta m_s=\pm2$ transition frequency [Fig. 2(c)] and does not correspond to any ODMR resonances (orange shaded stripes in Fig. 1(d)). To observe E-ODMR in the time domain, we fix the frequency of P to the $\Delta m_s=\pm2$ resonance and vary its length [Fig. 2(d)]. We observe electrically driven Rabi oscillations, whose frequency is modulated by the driving power [Fig. 2(e)]. The decay envelope is due to electric field inhomogeneity within the measurement volume and magnetic fluctuations of the coupled spin bath [9, 32] (See SI for data at other $B_\parallel$).

The frequency- and time-domain data in Fig. 2 indicate driving of the $\Delta m_s=\pm2$ transition. However, control measurements are necessary to confirm that population transfer was driven by a transverse electric field and not by stray transverse magnetic fields from the electrodes or driving circuit. Despite the fact that the $\Delta m_s=\pm2$ transition is magnetic-dipole forbidden, a misalignment of the nominally longitudinal magnetic field would result in a first-order mixing of the $|\pm1\rangle$ and $|0\rangle$ states and a second-order mixing of the $|-1\rangle$ and $|+1\rangle$ states. This mixing would then permit the nominal $\Delta m_s=\pm 2$ transition to be weakly driven by a transverse AC magnetic field. We rule out the magnetic driving scenario by performing the following controls.

We repeat the E-ODMR sequence, except that now we apply $P$ to the stripline, and attempt to drive the $\Delta m_s=\pm 2$ transition magnetically. To ensure that the transverse magnetic field is stronger in the control measurement [Fig. 3(a)] than in the E-ODMR measurement [Fig. 2(b)], we drive the stripline with 100 times more power than the electrodes had been driven (see SI). As expected for magnetic driving of the $\Delta m_s=\pm 2$ transition, a resonance is only seen when the longitudinal magnetic field is misaligned (orange curves in Fig. 3(b)). The same misalignment has little impact on the strength of the same resonance, when driven on the electrodes, as expected for electrical driving of the $\Delta m_s=\pm 2$ transition (blue curves in Fig. 3(b)).

We also observe that the E-ODMR Rabi frequency scales with the square root of the driving power, as expected [33] [Fig. 3(c)], and is independent of the $\Delta m_s=\pm 2$ resonance frequency. The latter point supports that the observed Rabi oscillations are driven electrically and not magnetically by displacement current through the interdigitated electrodes (see SI).

By spatially resolving the E-ODMR signal, we show that spin rotations are driven most efficiently within the interdigitated region. We fix $P$ to induce a π-pulse from $|-1\rangle$ to $|+1\rangle$ between the third and fourth electrode digits and spatially map $\Delta I_{PL}$ while raster-scanning the confocal excitation spot across the device [Fig. 4(a)-(c)]. The maximum of $\Delta I_{PL}$ is found to be in the interdigitated region, and Rabi driving inside and outside of this region [Fig. 4(d)] confirms that the decrease in $\Delta I_{PL}$ corresponds to a decrease in AC electric field strength. This result shows that high-frequency AC electric fields can be spatially mapped with the spin of optically addressable defects.

An important parameter that can be extracted from our time-domain measurements is the coupling strength between the ground-state spin and electric fields, $d_\perp$. For QL1 the electric Rabi frequency scales with square root of driving power as 0.97 MHz W$^{-1/2}$, from which we estimate that $d_\perp/h$=26 Hz cm V$^{-1}$. When extrapolated to the 2.4 MV cm$^{-1}$ dielectric strength of 6H-SiC [34], this figure implies that 60 MHz electrically driven Rabi oscillations should be possible with QL1. In contrast to the $g$-factor, which couples magnetic fields to the spin and is typically near 2 for isolated defects, $d_\perp$ is highly structure-and material-dependent [30]. Therefore, by appropriately selecting the material and defect, it might be possible to engineer the value of $d_\perp$ for a broad spectrum of applications.

We introduced and implemented E-ODMR to demonstrate that AC electric fields can be used to coherently control the spin of optically addressable defects. By spatially mapping the E-ODMR signal around our device, we demonstrated that this technique can confine spin control between electrodes and can be used to image high-frequency electric fields.

In the future, applying E-ODMR to nanoscale devices could lead to the individual control and addressability of strongly-coupled spins in a dense network. In hybrid systems, E-ODMR could be used for transduction between spin and other quantum mechanical degrees of freedom. Other applications could use the fact that, together, magnetic and electric fields can drive arbitrary transitions across all three spin-triplet sublevels, a promising strategy for streamlining quantum-information algorithms [35]. Electrically driven spin resonance in optically addressable defects opens many exciting prospects for scalable quantum control.

Acknowledgment

We acknowledge helpful conversations with Will Koehl, David Toyli, and Joe Heremans. This work was supported by AFOSR, DARPA, NSF.

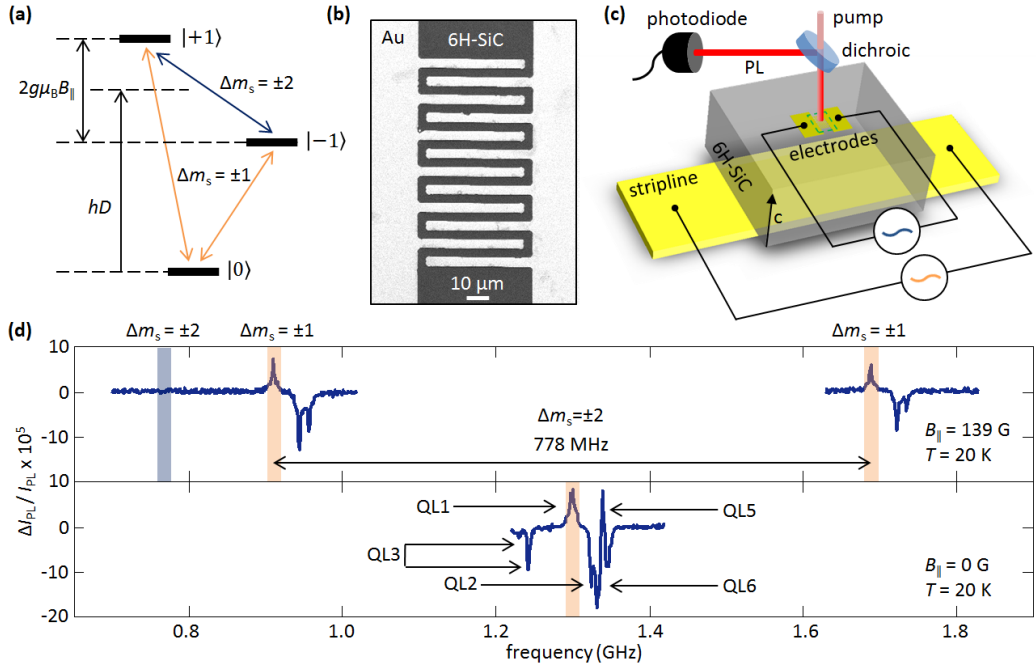

FIG. 1. (a). The orbital ground-state spin structure of the QL1 defect, with $\Delta m_s=\pm 1$ transitions (orange arrows) and the $\Delta m_s=\pm 2$ transition (blue arrow) indicated. (b) Scanning electron microscope image of the electrode pattern. (c) QL1 spins localized within a 400 nm thick layer immediately beneath the 6H-SiC surface are optically pumped with a 1.27 eV laser in a 1.5 μm diameter spot, addressing ~$10^4$ QL1 defects at once. Spins are driven electrically by the electrodes and magnetically by the stripline. The electrode pattern from part (b) maps to the green dashed parallelogram. (d) ODMR signal when the stripline is driven at $B_\parallel$=139 G (upper) and $B_\parallel$=0 G (lower) at $T$=20 K. The $\Delta m_s=\pm 1$ transitions are shaded orange and the $\Delta m_s=\pm 2$ transition (at 778 MHz, not seen in ODMR) is shaded blue.

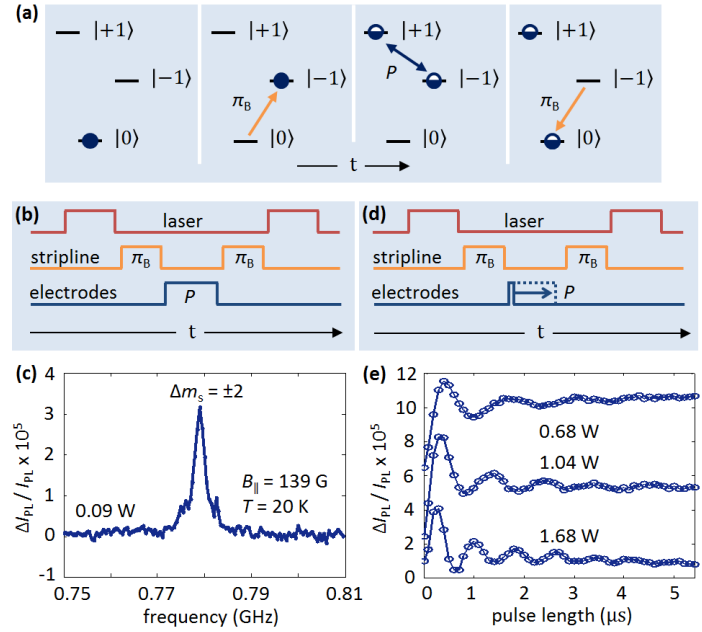

FIG. 2. (a) The sequence of spin transitions used to observe E-ODMR. (b) Pulse sequence used for the frequency-domain E-ODMR measurement. The width of *P* is fixed and its frequency is swept. (c) A clear E-ODMR feature is seen at the frequency difference of the $|+1\rangle$ and $|-1\rangle$ states, indicating population transfer across the $\Delta m_s = \pm 2$ transition. (d) Pulse sequence used for time-domain E-ODMR measurements. The frequency of *P* is fixed to the $\Delta m_s = \pm 2$ resonance and its length is varied, resulting in (e), Rabi oscillations at three separate electrode driving powers.

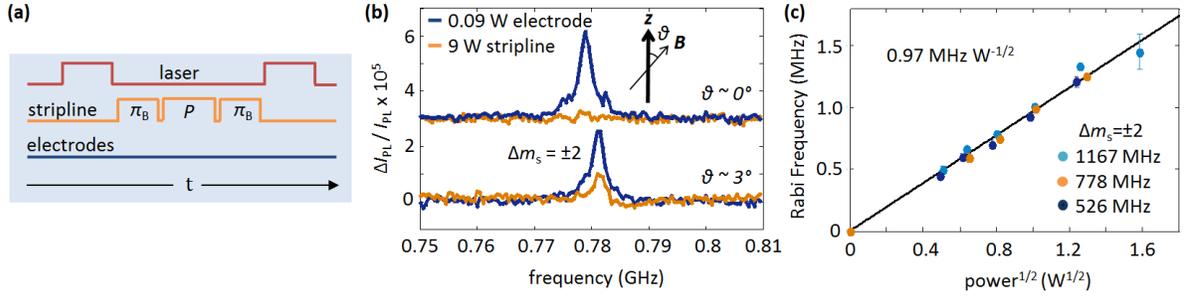

FIG. 3. (a) Pulse sequence used for the control measurement in which we attempt to drive the $\Delta m_s=\pm 2$ transition magnetically with the stripline. The width of $P$ is fixed and its frequency is swept. (b) A feature using the pulse sequence in (a) is only observed at the $\Delta m_s=\pm 2$ resonance when the static longitudinal magnetic field is purposefully misaligned to the z-axis (orange curves). However, E-ODMR (blue curves) results in a $\Delta I_{PL}$ feature at the $\Delta m_s=\pm 2$ resonance regardless of the magnetic field misalignment. (c) The electrically driven Rabi frequency scales with the square root of the driving power (0.92 MHz W$^{-1/2}$) and is independent of the $\Delta m_s=\pm 2$ frequency. The error bars represent the 95% confidence intervals from fits to Rabi curves.

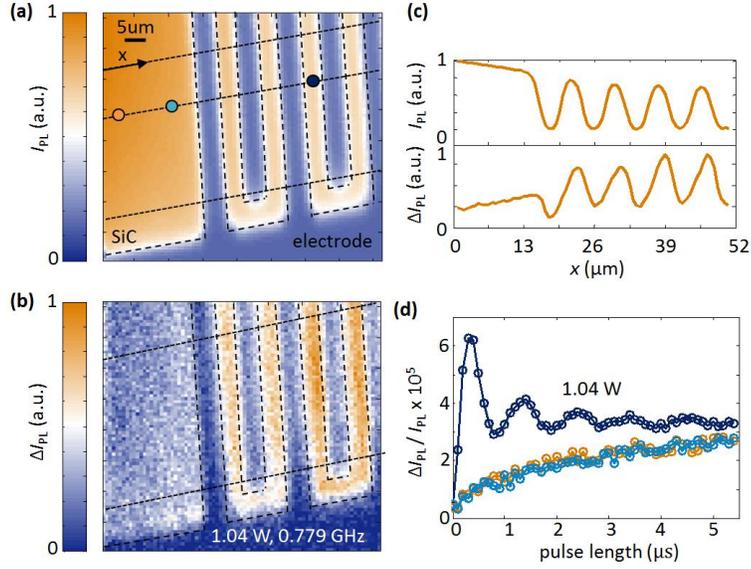

FIG. 4. (a) A map of $I_{PL}$ in a region of our electrode device. The electrodes are blue. (b) $\Delta I_{PL}$ in the same region, with the length and frequency of $P$ set to induce a full spin rotation from $|-1\rangle$ to $|+1\rangle$ between the third and fourth electrode digits (dark blue circle in (a)). The maximum of $\Delta I_{PL}$ is measured in the interdigitated region, indicating that spin rotations are spatially confined by the electrodes. (c) $I_{PL}$ (upper frame) and $\Delta I_{PL}$ (lower frame) averaged along the length of the electrode digits between the dashed black lines in (a) and (b). $x$ is defined to be position along the dashed lines. (d) Rabi curves taken at the three color-coded circles indicated in (a). Outside of the electrodes, the transverse electric field is too weak to drive Rabi oscillations.
11